\documentclass[12pt]{article}
\usepackage{amsfonts}
\usepackage{epsfig}
\usepackage{url}
\usepackage{color}
\usepackage{amsfonts}
\usepackage{mathptmx}
\usepackage{helvet}
\usepackage{courier}
\usepackage{type1cm}         

\begin{document}

\title{ Urban Landscape is an Important Factor in Rising Inequality, Spatial Segregation, and Social Isolation}

\author{\noindent\large  Philippe Blanchard, Dimitri Volchenkov  \footnote{Corresponding author. E-mail address:volchenk@physik.uni-bielefeld.de} }

 \maketitle

\begin{center}
 {\it   Faculty of Physics,  Bielefeld University, Universitaetsstr. 25, 33615 Bielefeld, Germany} 
\end{center}

\begin{abstract}

Urbanization has been the dominant demographic trend
 in the entire world, during the last half century.
Rural to urban migration, international migration,
 and the re-classification or expansion of existing
city boundaries  have been among
the major reasons for increasing urban population.
The essentially fast growth of cities
 in the last decades
urgently calls
for a profound insight into
the common principles
stirring the structure of  urban developments
all over the world.
 
We  have discussed
the graph representations of urban spatial structures
and suggested a
computationally simple technique 
that can be used
in order to  spot
the relatively  isolated locations and neighborhoods,
to detect urban sprawl, and to illuminate
 the hidden community structures in
complex urban textures.
The approach  may be implemented for
  the detailed expertise
of any urban pattern and
the associated transport networks
that may include many transportation modes.

\end{abstract}

\begin{flushleft}
\end{flushleft}

\renewcommand{\baselinestretch}{1.3}
\normalsize

\tableofcontents

\section{Introduction \label{Sec:Intro}}
\noindent
 
A belief in the influence of the built environment on humans was common in architectural and urban thinking for centuries. Cities generate more interactions with more people than rural areas because they are central places of trade that benefit those who live there. People moved to cities because they intuitively perceived the advantages of urban life. City residence brought freedom from customary rural obligations to lord, community, or state and converted a compact space pattern into a pattern of relationships
by constraining mutual proximity between people.
Spatial organization
of a place
has an extremely important effect
on the way
people move through spaces
and meet other people by chance, \cite{Hillier:1984}.
Compact neighborhoods can foster casual
social interactions among neighbors,
while creating barriers to interaction
 with people outside a neighborhood.
Spatial configuration promotes peoples encounters as well as
making it possible for them to avoid each other, shaping social
patterns, \cite{Ortega-Andeane:2005}.

The phenomenon
of   {clustering of minorities},
especially that of newly  arrived immigrants,
is well documented
since the work of \cite{Wirth:1928}
(the reference appears
 in \cite{Vaughan:2005a}).
Clustering is considering
to be beneficial for mutual support and
for the sustenance of cultural
and religious activities.
At the same time,
clustering and the subsequent physical segregation of
minority groups would cause their   {economic marginalization}.
The study of London's change
over 100 years performed by \cite{Vaughan:2005b}
has indicated
that the creation of poverty areas
is a spatial process:
by looking at the distribution of poverty at the street,
it is possible
to find a relationship
between spatial segregation and
poverty.
The   {patterns of mortality} in London studied over the past
century by \cite{Orford:2002}
show that the areas
of persistence of
poverty
cannot be explained other
than by an underlying spatial effect.

Urban planning is recognized to play a crucial position in the development of sustainable cities.  The essentially fast growth of cities in the last decades urgently calls for a profound insight into the common principles stirring the structure of urban development  all over the world.

Sociologists think that isolation worsens an area's economic prospects by reducing opportunities for commerce, and engenders a sense of isolation in inhabitants, both of which can fuel poverty and crime. Urban planners and governments have often failed to take such isolation into account when shaping the city landscape, not least because isolation can sometimes be difficult to quantify in the complex fabric of a major city. 
The source of such a difficulty is profound:
 while humans live and act in Euclidean space 
which they percept visually as affine
space and which is present in them as a mental form, 
a complex network of interconnected spaces of movements 
that constitutes a spatial urban pattern 
does not possess the structure  of Euclidean space.
 In another circumstance we
spoke of fishes: they know nothing either of what the sea, or a lake, or a river might really be and only know fluid as if it were air around them. While in a complex built environment, humans have no sensation of it, but need time to construct its "affine representation" so they can understand and store it in their spatial memory. Therefore, human behaviors in complex environments result from a long learning process and the planning of movements within them.
In \cite{Blanchard:2009}, we suggested that random walks can help us to find such an "affine representation" of the built environment, giving us a leap outside our Euclidean "aquatic surface" and opening up and granting us the sensation of new space.

While travelling in the city, our primary interest is often in
finding the best route from one place to another. Since the
  {way-finding} process is a purposive, directed, and motivated
activity \cite{Golledge:1999}, the shortest route
 is not necessary the best one.
If an origin and a destination
are not directly connected
by a continuous path,
wayfinding may include
 search and exploration actions
for which
it may be crucial to recognize juxtaposed
 and distant
 {landmarks}, to determine turn angles
and directions of movement,
and eventually  to embed
the route in some
large reference frame. It is well known
 that the conceptual representations of
space in humans do not bear a one-to-one correspondence with
actual physical space. The process of integration of the
local affine models of individual places
 into
the entire cognitive map of the urban area network
  is very
complicated and falls largely within the domain of
cognitive science and psychology,
but nevertheless
the specification of what
may be recovered from
 spatial memory can be considered as a problem of mathematics --
"the limits of human perception coincide with
 mathematically plausible solutions",
\cite{Pollick:1997b}. Supposing the inherent mobility of humans and alikeness
 of their spatial perception aptitudes,
one might
argue that
nearly all people experiencing the city
would agree
in their judgments on the total number of
individual locations in that,
in identification of the borders of these locations,
 and their interconnections.
In other words, we assume that
spatial experience in humans
intervening in the city
may be organized in the form of
a universally acceptable network.

Well-known and frequently travelled path segments
 provide   {linear anchors} for certain
city districts and neighborhoods that helps to
organize collections of
spatial models for the
individual locations into
a configuration representing
the mental image of the entire city.
In our study, we assume that
 the frequently travelled routes
 are nothing else
but the "{projective invariants}" of the given layout of streets
and squares in the city -- the function of its geometrical
configuration, which remains invariant whatever origin-destination
route is considered. The arbitrary linear transformations of the
geometrical configuration with respect to
 which
a certain property remains invariant
constitute the generalized affine transformations.

It is intuitively clear  that
 if the spatial configuration
 of the city is represented by a regular graph,
 where each location represented by
 a vertex has
 the same number of neighbors,
in absence of other local landmarks,
all paths would be equally probably
followed by travelers.
No linear anchors are possible in such an urban pattern
which
could stimulate spatial apprehension.
However, if the spatial graph of the city
is far from being regular,
then a configuration disparity of
different places in the city
 would result in that some of them
may be visited by travelers more often than others.

In the following sections of our work, 
we study the problem of isolation in cities
with the use of random walks that provide
us with an effective tool for the
detailed structural analysis
of connected undirected graphs exposing their
symmetries, \cite{Blanchard:2009}.

\section{Spatial graphs of urban environments} 
\label{sec:Networks}
\noindent

In traditional
 urban researches, the dynamics of an urban pattern come
 from the landmasses, the physical aggregates of buildings
 delivering place for people and their activity.
The relationships between certain components of the urban texture
  are often measured along streets and routes considered as edges
  of a planar graph, while the traffic end points and street
  junctions are treated as nodes. Such a   primary graph
  representation of urban networks is grounded on relations
  between junctions through the segments of streets. The usual
   city map based on Euclidean geometry can be considered as an
   example of primary city graphs.

In space syntax theory (see \cite{Hillier:1984,Hillier:1999}),
 built environments are treated as systems
of spaces of vision subjected to a configuration analysis.
Being
irrelevant to the physical distances, spatial graphs
 representing the
urban environments are
removed from the physical space.
It has been demonstrated in multiple experiments
that spatial perception
shapes peoples understanding of how
a place is organized and eventually  determines the pattern of local
 movement, \cite{Hillier:1999}.
The aim
of the space syntax  study is to estimate the relative proximity
between different locations and to associate these distances to
 the densities of
human activity along the links connecting them,
\cite{Hansen:1959,Wilson:1970,Batty:2004}. The surprising accuracy
of predictions of human behavior in cities based on the purely
topological analysis of different urban street layouts within the
space syntax approach attracts meticulous attention
\cite{Penn:2001}.

The decomposition of
urban spatial networks
into the complete sets
of intersecting open spaces
can be based on a number of different principles.
In  \cite{Jiang:2004},
while identifying a street over a plurality of routes
on a city map, the  named-street approach has been used, in
which two different arcs of the primary city network were
assigned to the same identification number (ID) provided they share the same
street name.

In the present chapter,
 we take a "named-streets"-oriented point of view
on the decomposition of
urban spatial networks
into the complete sets
of intersecting open spaces
 following our previous works \cite{Volchenkov:2007a,Volchenkov:2007b}.
 Being interested in the statistics of random walks defined on spatial
networks of urban patterns, we assign an individual
street ID code to each continuous segment of a street. The spatial
 graph of urban environment is then constructed by
mapping all edges (segments of streets) of the city map
shared the same street ID into nodes
 and all intersections among each pair of edges of the primary graph
into the edges of the secondary graph connecting the corresponding nodes.

Although graphs are usually shown diagrammatically,
they can also be represented as matrices.
The major
advantage of matrix representation
 is that then the analysis of graph structure
 can  be performed using well known operations on
  matrices.
For each graph, there exists a unique  {adjacency matrix} (up to
permuting rows and columns) which is not the adjacency matrix of
any other graph. 
If we assume that the  spatial graph of the city is simple
(i.e., it contains neither loops, nor multiple edges), the
adjacency matrix is a $\{0,1\}$-matrix with zeros on its diagonal:
\begin{equation}
\label{adjacencymatrix}
A_{ij}\,=\,\left\{
\begin{array}{ll}
1,& i\sim j, \quad i\ne j, \\
0,& \mathrm{otherwise}.
\end{array}
\right.
\end{equation}
If the graph is undirected, the adjacency matrix is symmetric, $A_{ij}=A_{ji}$.
If the graph contains  twins nodes,
the correspondent rows and columns of $\bf A$
  are identical.

\section{Graphs as discrete time dynamical systems}
\label{sec:Ruelle-Perron-Frobenius_operator_of_a_graph}
\noindent

A finite connected undirected graph  can be seen as a 
{\it discrete time dynamical system} possessing a finite number
of states (nodes) \cite{Prisner:1995}.
The behavior of such a dynamical
system can be studied by means of a
  transfer operator  
 which describes 
the time evolution of
distributions in phase space. 
The transfer operator can be represented  
by a stochastic matrix 
determining 
a discrete time random walk 
 on the graph
in which a  
walker
picks 
at each node
between
 the various available edges
 with equal probability.
An obvious benefit 
of the approach
based on random walks
to graph theory
 is that  
the relations between individual nodes 
and subgraphs 
acquire a 
precise quantitative 
probabilistic 
description 
that enables us to attack applied problems 
which could not even be started otherwise.
Given a finite connected undirected graph 
$G(V,E),$ let us 
consider a  transformation 
$
\mathcal{S}:\,V\,\to \,V
$ 
mapping
 any subset of nodes $U\subset V$ into the set of
their 
direct 
neighbors,
$\mathcal{S}(U)=\left\{w\in V|v \in U,v\sim w\right\}.
$
We denote the result 
of $t\geq  1$ consequent applications 
of  
$\mathcal{S}$
to $U\subset V$ as $\mathcal{S}_t(U)$.
The iteration of the map $\mathcal{S}$ 
leads to a study of
possible 
paths 
in the graph $G$
beginning at $v\in V.$
However,
 we rather discuss 
the time evolution of smooth functions
  under iteration,
than the individual 
trajectories $\mathcal{S}_t(v).$
Given a discrete {\it density function}
 $ 
f(v)\geq 0, $ $ v\in V,
$
defined on  a undirected connected graph
$G(V,E)$
 such that
$
\sum_{v\in V}f(v)=1,
$
the dynamics 
of the map
  $\mathcal{S}(U)$
is described by the norm-preserving 
transformation
\begin{equation}
\label{PerronFrobenius}
\sum_{v\,\in \,U }\,\,f(v)\,{\bf T}^{t} 
\,\,=\,\,\sum_{\mathcal{S}^{-1}_t(U)}\, f(v),
\end{equation}
where 
 ${\bf T}^{t}$ is the  
   {\it Ruelle - Perron - Frobenius} {\it transfer operator}
  corresponding to the transformation
 $\mathcal{S}_t.$
The uniqueness of the Ruelle - Perron - Frobenius 
operator  for a
given transformation
 $\mathcal{S}_t$ is a
consequence of the 
   Radon - Nikodym theorem 
extending the concept 
of  probability densities
to probability measures 
defined over arbitrary sets, \cite{Shilov:1978}.
It was shown by \cite{Mackey:1991} 
 that the relation (\ref{PerronFrobenius})
is satisfied by
a homogeneous
 Markov chain $\{v_t\}_{t\in \mathbb{N}}$
determining 
a  random walk
of the nearest neighbor type
defined on the connected undirected 
graph $G(V,E)$
by the transition matrix
\begin{equation}
\label{T_rw}
\begin{array}{lcl}
T_{ij} & =& \Pr \left[\left.v_{t+1}=j\right|v_t=i\right] >0 
\Leftrightarrow \,\, i\sim j,\\
       & =& \mathbf{D}^{-1}\mathbf{A}, \quad
\mathbf{D}=\mathrm{diag}(\deg(1),\ldots \deg(N)), \quad
\deg(i)\equiv\sum_{j=1}^NA_{ij}.
\end{array}
\end{equation}
where $\bf A$ is the adjacency matrix of the graph,
so that 
the probability of
 transition  from $i$ to $j$ in $t>0$ steps
equals
$
p^{(t)}_{ij}=\left({\bf T}^t\right)_{ij}.
$
 The discrete time random
walks on graphs have been 
studied in details in
\cite{Lovasz:1993,Lovasz:1995} and 
by many other authors. 
For a random walk defined on a 
connected undirected graph,
the Perron - Frobenius theorem 
  asserts the unique 
strictly positive
probability
 vector 
$ 
{\bf\pi}  =  \left(\pi_1,\ldots,\pi_N\right)
$ 
(the left eigenvector of the transition matrix 
${\bf T}$ belonging to the maximal eigenvalue $\mu=1$)
 such that 
$
{\bf \pi}{\bf T}=1\cdot{\bf\pi}.
$ 
For the
nearest neighbor random walks defined
on an undirected graph, 
the {\it stationary
distribution} of random walks 
 on an undirected graph
equals
\begin{equation}
\label{stat-distribution}
\pi_i\,=\,\frac{\,\,\deg(i)\,\,}{\,\,2|E|\,\,},\quad \sum_{i\in V} \pi_i\,=\,1
\end{equation}
where $\deg(i)$ is the number of immediate neighbors of the node $i$, and $|E|$ is the total number of edges in the graph. 
The vector ${\bf \pi}$ satisfies 
the condition of detailed   
 balance, 
$\pi_i\,T_{ij} = \pi_j\,T_{ji},
$ from which it follows that a random walk
defined on an undirected graph 
is time reversible: 
it is
also a random walk if 
considered backward, and it is not possible 
to determine, given the walker  at a number
 of nodes in time after running
 the 
walk, which state came first and which state arrived later.
The stationary distribution (\ref{stat-distribution})
of random walks defined on a connected undirected 
graph $G(V,E)$
determines a unique measure on $V,$
$D=\sum_{j\in V}\deg(j)\delta_j,
$
with respect to which the 
 transition operator (\ref{T_rw})
becomes self-adjoint and is
represented by a symmetric transition matrix,
\begin{equation}
\label{T_symm}
\widehat{T_{ij}}=\left({\bf D}^{1/2}\,{\bf T}\,{\bf D}^{-1/2}\right)_{ij} 
                = \frac {\,\,A_{ij}\,\,}{\,\,\sqrt{\deg(i)\deg(j)}\,\,}
\end{equation}
where $\bf D$ is the diagonal matrix of graph's degrees.

Diagonalizing the symmetric matrix (\ref{T_symm}),
we obtain 
$
\widehat{\bf T}=\mathrm{\Psi}{\bf M}\mathrm{\Psi}^\top,
$
where $\mathrm{\Psi}$ is an orthonormal matrix,
$
\mathrm{\Psi}^{\top}=\mathrm{\Psi}^{-1},
$
 and $\bf M $ is a
 diagonal matrix with entries 
$1=\mu_1>\mu_2\geq\ldots\geq\mu_N> -1
$ (here,
 we do not consider bipartite graphs,
 for which $\mu_N=-1$).
The rows 
${\bf \psi}_k=\left\{
\psi_{k,1},\ldots,\psi_{k,N}
\right\}$
 of the orthonormal matrix $\mathrm{\Psi}$
forms
an orthonormal
 basis
in Hilbert space $\mathcal{H}(V),$
$ {\bf \psi}_k:\,V\to S_1^{N-1},$ $k\,=\,1,\ldots N,
$ where $S_1^{N-1}$ is the $N-1$-dimensional unit sphere.
We consider 
the eigenvectors $ {\bf \psi}_k$ ordered 
in accordance to the  eigenvalues they belong to.
For eigenvalues of  algebraic multiplicity $\alpha>1$,
 a number of
linearly independent 
orthonormal
 ordered
 eigenvectors can be chosen to span
the associated eigenspace.
The first eigenvector
    $ {\bf \psi}_1$ belonging to
 the largest eigenvalue $\mu_1=1$
(which is simple)
is the Perron-Frobenius eigenvector 
that determines the
stationary distribution of random walks
over the graph nodes, $ \psi_{1,i}^2\,=\,\pi_i$, 
$i=1,\ldots,N$.
 The squared 
Euclidean norm of the vector
in the orthogonal complement of $\vec{\psi}_1$,
$
\sum_{s=2}^N\psi_{s,i}^2 = 1-\pi_i>0$, 
expresses
the probability
that a random walker is not  in $i$.

\section{Affine probabilistic geometry of eigenvector embedding  and probabilistic Euclidean distance on graphs}
\label{sec:FPT-embedding}
\noindent

Discovering of important nodes
and quantifying differences between them
 in a graph
is not easy, since
the graph
does not possess {\it a priori}
the structure of Euclidean space.
We 
 use 
  the algebraic properties 
of the 
self-adjoint  operators
in order to define 
an Euclidean metric
on any finite connected undirected graph.
Geometric objects,
 such as points, lines, or planes, can be given
a representation as elements in projective space based on
{\it homogeneous coordinates}.
Given an orthonormal basis
 $\left\{ {\bf \psi}_k:V\to S_1^{N-1}\right\}_{k=1}^N$
in $\mathbb{R}^N,$
any vector in
Euclidean space 
can be expanded into
$
\mathbf{v}=
\sum_{k=1}^N\left\langle\mathbf{v}|\psi_k\right\rangle\left\langle\psi_k\right|.
$ Provided $\left\{\vec{\psi}_k\right\}_{k=1}^N$
are the eigenvectors of the symmetric 
matrix of the 
operator $\widehat{\bf T},$
we can define
new basis vectors,
$\mathrm{\Psi}'\equiv \left\{1,\frac{\psi_{2,2}}{\psi_{1,2}},\ldots,
\frac{\psi_{N,N}}{\psi_{1,N}}
\right\}, 
$ since we have always $\psi_{1,i}\equiv\sqrt{\pi_i}>0$ for any
$i\in V.$
The new basis vectors 
 span 
the projective space $P\mathbb{R}_{\pi}^{(N-1)},$
so that the vector ${\bf v}$ 
 can be expanded into  
\begin{equation}
\label{expansion_Euclidean_2}
\mathbf{v}{\bf \pi}^{-1/2}\,=\,
\sum_{k=2}^N\left\langle {v}|\psi'_k\right\rangle\left\langle
\psi'_k\right|.
\end{equation}
It is easy to see 
 that
the transformation (\ref{expansion_Euclidean_2})
defines a stereographic projection 
on $P\mathbb{R}_{\pi}^{(N-1)}$
such that 
all vectors in $\mathbb{R}^N(V)$
collinear to the 
vector  $\mid\psi_1\rangle$
corresponding to the 
stationary distribution
of random walks
 are projected onto a common image point.
If the graph $G(V,E)$ has some isolated nodes
$\iota\in V,$ 
for which $\pi_{\iota}=0,$
they 
 play the role of the   plane at
infinity  with respect to (\ref{expansion_Euclidean_2}),
 away from which
we can use the basis  $\mathrm{\Psi}'$ as an ordinary Cartesian system.
  The
transition to the homogeneous coordinates  
transforms vectors of $\mathbb{R}^N$ into vectors on the
$(N-1)$-dimensional hyper-surface $\left\{
\psi_{1,x}=\sqrt{\pi_x}\right\}$, the orthogonal
complement to the
 vector of
stationary distribution $ {\bf \pi}$.

The Green function (a pseudo-inverse) of the normalized Laplace operator
$\widehat{\bf L}={\bf 1}-\widehat{\bf T}$ describing the diffusion of 
random walkers over the undirected graph 
is given in the 
homogeneous coordinates  by
$
\widehat{\bf L}^\natural = \sum_{k=2}^N\,
 \lambda_k^{-1} { \left|\psi'_k\right\rangle\left\langle\psi'_k\right| }.
$
In order to
obtain a Euclidean metric on the
 graph $G(V,E)$,
one needs to introduce
  distances between points (nodes of the graph)
 and the angles between vectors pointing at them
that can be done by determining the 
dot product between
 any two vectors $\vec{\xi},\vec{\zeta} \in P\mathbb{R}_{\pi}^{(N-1)}$
by
$
\left(\vec{\xi},\vec{\zeta}\right)_{T} = 
 \left(\vec{\xi},{\bf L}^\natural\vec{\zeta}\right).
$
The dot product  
is a symmetric real valued scalar function that
allows us to define the (squared)
norm of a vector $\vec{\xi}\in P\mathbb{R}_{\pi}^{(N-1)}$  by
$
\left\|\, \vec{\xi}\,\right\|^2_{T} = 
\left(\vec{\xi},{\bf L}^\natural\vec{\xi}\right).
$
The  squared 
norm 
\begin{equation}
\label{norm_node}
\left\|\,\mathbf{e}_i\,\right\|_T^2\, =\,\frac 1{\pi_i}\,\sum_{s=2}^N\,
\frac{\,\psi^2_{s,i}\,}{\,\lambda_s\,}.
\end{equation}
of 
the canonical basis vector 
$\mathbf{e}_i=\{0,\ldots 1_i,\ldots 0\}$ 
representing
the node $i\in V$ is nothing else, but 
the spectral representation of the   {\it first passage time}
to the node $i\in V$,
the expected number of steps required to
reach the node $i\in V$ for the first time
starting from a node randomly chosen
among all nodes of the graph
accordingly to the stationary distribution $\pi$.
The first passage time, $\left\|\mathbf{e}_i\right\|_T^2$,
 can be directly used in order to characterize
the level of accessibility of the node $i$, \cite{Blanchard:2011}.
The Euclidean distance between two vectors, $\mathbf{e}_i$ 
and $\mathbf{e}_j$, given by 
$
\left\|
   \mathbf{e}_i- \mathbf{e}_j\right\|^2_{T}  =  \left\|\, \mathbf{e}_i\,\right\|^2_{T}+\left\|\, \mathbf{e}_j\,\right\|^2_{T}-2\left(\mathbf{e}_i,\mathbf{e}_j\right)_{T}
$
is nothing else, but 
 the   {\it commute time}, the expected number of
steps required for a random walker starting at $i\,\in\, V$ to
visit $j\,\in\, V$ and then to return back to $i$, \cite{Blanchard:2011}.

\section{First-passage times to ghettos}
\label{subsec:FPT_Ghettos}
\noindent

The phenomenon of clustering of minorities, especially that of newly
 arrived immigrants, is well documented \cite{Wirth}.
Clustering is considering to be beneficial for mutual support and
for the sustenance of cultural and religious activities. At the
same time, clustering and the subsequent physical segregation of
minority groups would cause their economic marginalization.
The spatial analysis of the immigrant
quarters \cite{Vaughan01}  and the
study of London's changes over 100 years \cite{Vaughan02}
shows that they were significantly more
segregated from the neighboring areas, in particular, the number
of street turning away from the quarters to the city centers were
found to be less than in the other inner-city areas being
usually socially barricaded by railways, canals and industries.
It has been suggested \cite{Language} that space structure and its
impact on  movement are critical to the link between the built
environment and its social functioning. Spatial structures creating a local situation
in which there is no relation between movements inside the spatial
pattern and outside it and the lack of natural space occupancy
become associated with the social misuse of the structurally
abandoned spaces.

\begin{figure}[ht]
 \noindent
\epsfig{file=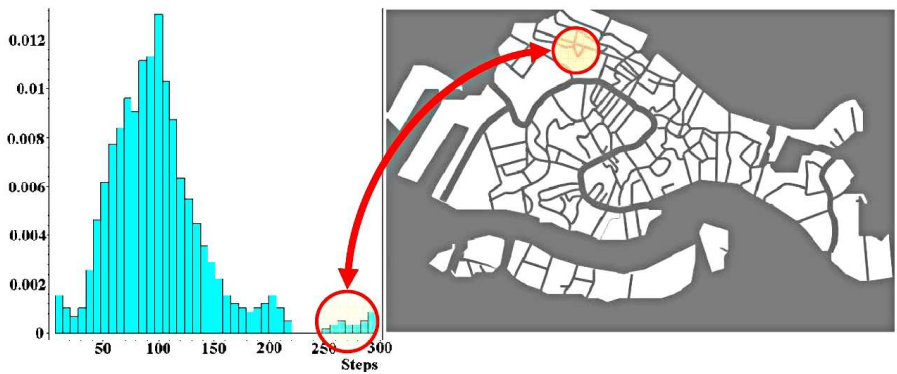,  angle= 0,width =12cm, height =5cm}
\caption{\small  The Venetian Ghetto jumped out
as by far the most isolated, despite being apparently well connected to the rest of the city. }
 \label{Fig3_urban}
\end{figure}

We have analyzed the first-passage times to individual canals in the
spatial graph of the canal network in Venice.
The distribution of numbers of canals
over the range of the first--passage time values is represented
by a histogram shown in Fig.~\ref{Fig3_urban}.left.
 The height of each bar in the histogram
is proportional to the number of canals in the
 canal network of Venice for which the first--passage
times fall into the disjoint intervals (known as bins).
Not surprisingly,
the Grand Canal, the giant Giudecca Canal
 and the Venetian lagoon are the most connected.
In contrast,  the Venetian Ghetto (see Fig.~\ref{Fig3_urban}.right) -- jumped out
as by far the most isolated, despite being apparently well connected to the rest of the city --
 on average, it took 300 random steps to reach, far more than the average of 100 steps for other places in Venice.

The Ghetto was created in March 1516 to separate Jews from the Christian majority of Venice. It persisted until
1797, when Napoleon conquered the city and demolished the Ghetto's gates.
Now it is abandoned.

\section{Why is Manhattan so expensive?}
\label{sec:manhattan}
\noindent 

The notion of
isolation
 acquires
the statistical interpretation by means of random walks. The
first-passage times in the city vary strongly from  location to
location. Those places characterized by the shortest first-passage
times are easy to reach while very many random steps would be
required in order to get into a statistically isolated site.

Being a global characteristic
of a node in the graph,
the first-passage time
assigns  absolute scores
to all nodes
 based on the probability
 of paths they provide
 for random walkers.
The first-passage time
can therefore be considered
as a natural
statistical
centrality measure of
 the node within the graph, \cite{Blanchard:2009}.

 \begin{figure}[ht]
 \noindent
\begin{center}
\epsfig{file=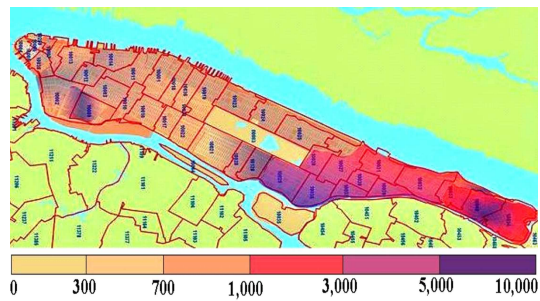, angle= 0,width =7cm, height =4cm}
  \end{center}
\caption{\small Isolation map of Manhattan. Isolation is measured
by  first-passage times to the places.  Darker color corresponds
to longer first-passage times.}
 \label{Fig2_Isolation}
\end{figure}

A  visual pattern
displayed on Fig.~\ref{Fig2_Isolation}
represents the pattern of structural
  isolation (quantified by the first-passage times)
 in Manhattan (darker color corresponds to longer first-passage times).
It is interesting to note that the  {spatial distribution of
isolation} in the urban pattern of Manhattan
(Fig.~\ref{Fig2_Isolation})
 shows a qualitative agreement with the map
of the  tax assessment  value of the land in Manhattan reported by
B. Rankin (2006) in the framework of the RADICAL CARTOGRAPHY
project being practically a negative image of that.

\begin{figure}[ht!]
 \noindent
\centering
\begin{tabular}{llll}
1). & \epsfig{file= 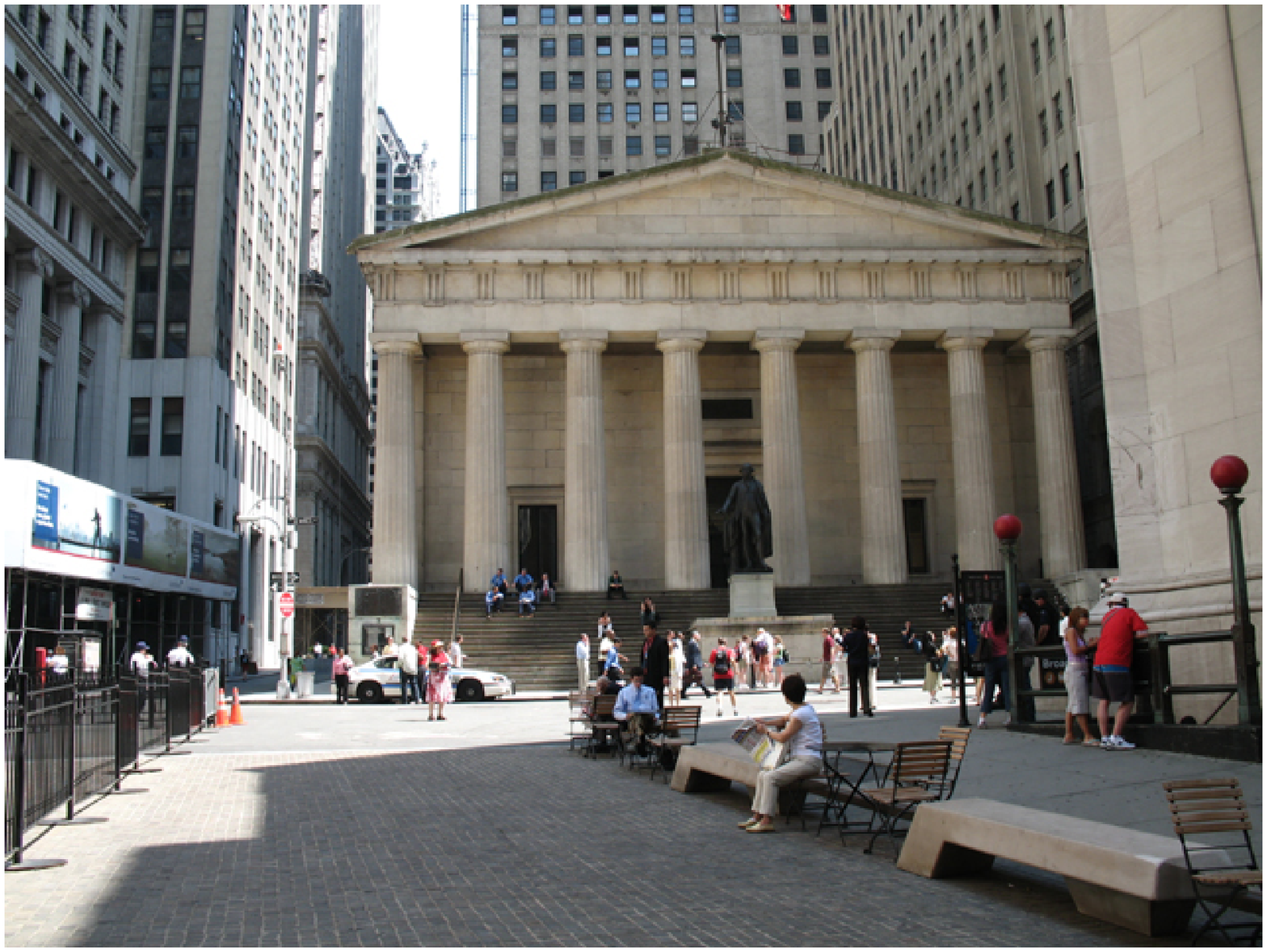, width=6cm, height =4cm}
& 2). &  \epsfig{file= 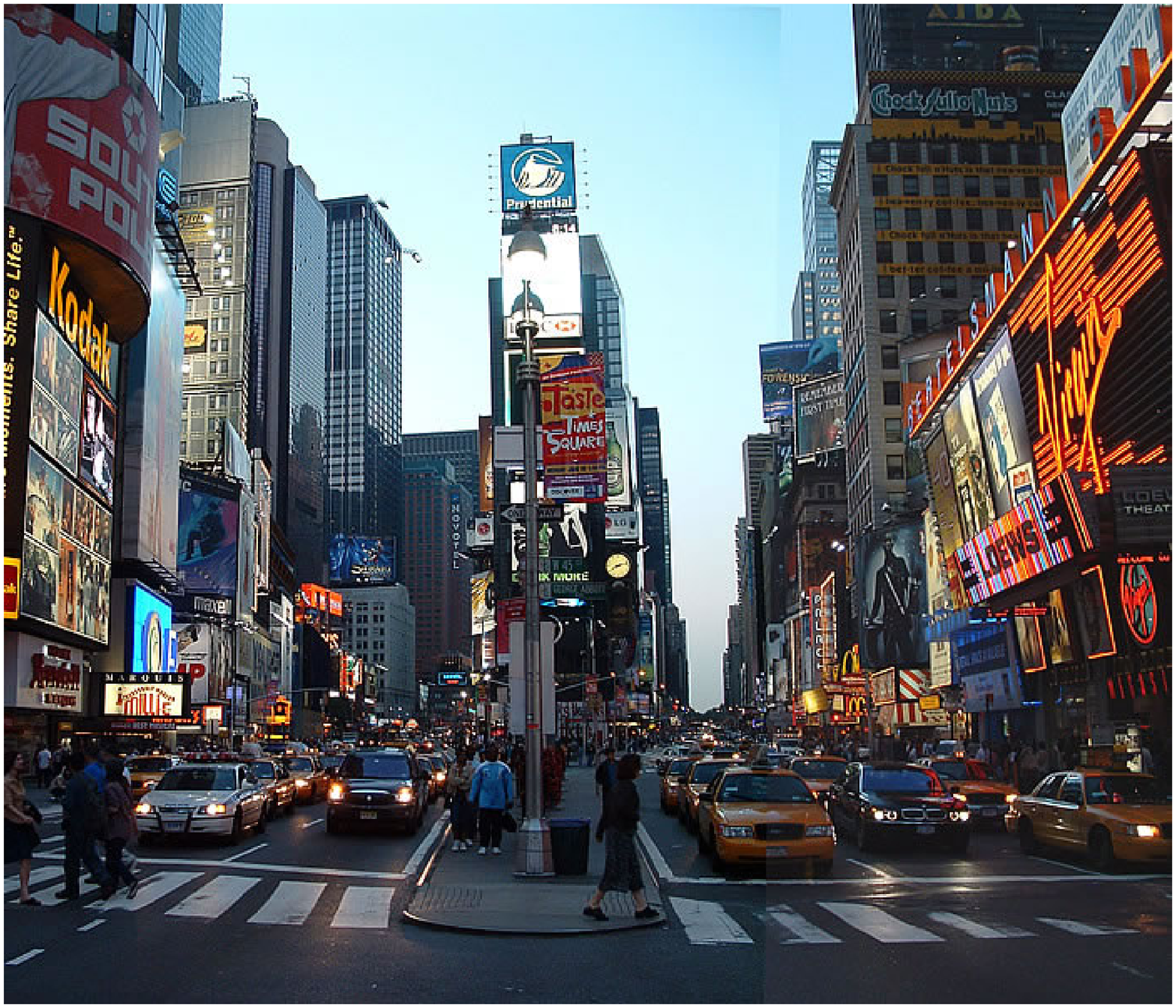, width=6cm, height =4cm} \\
3). & \epsfig{file= 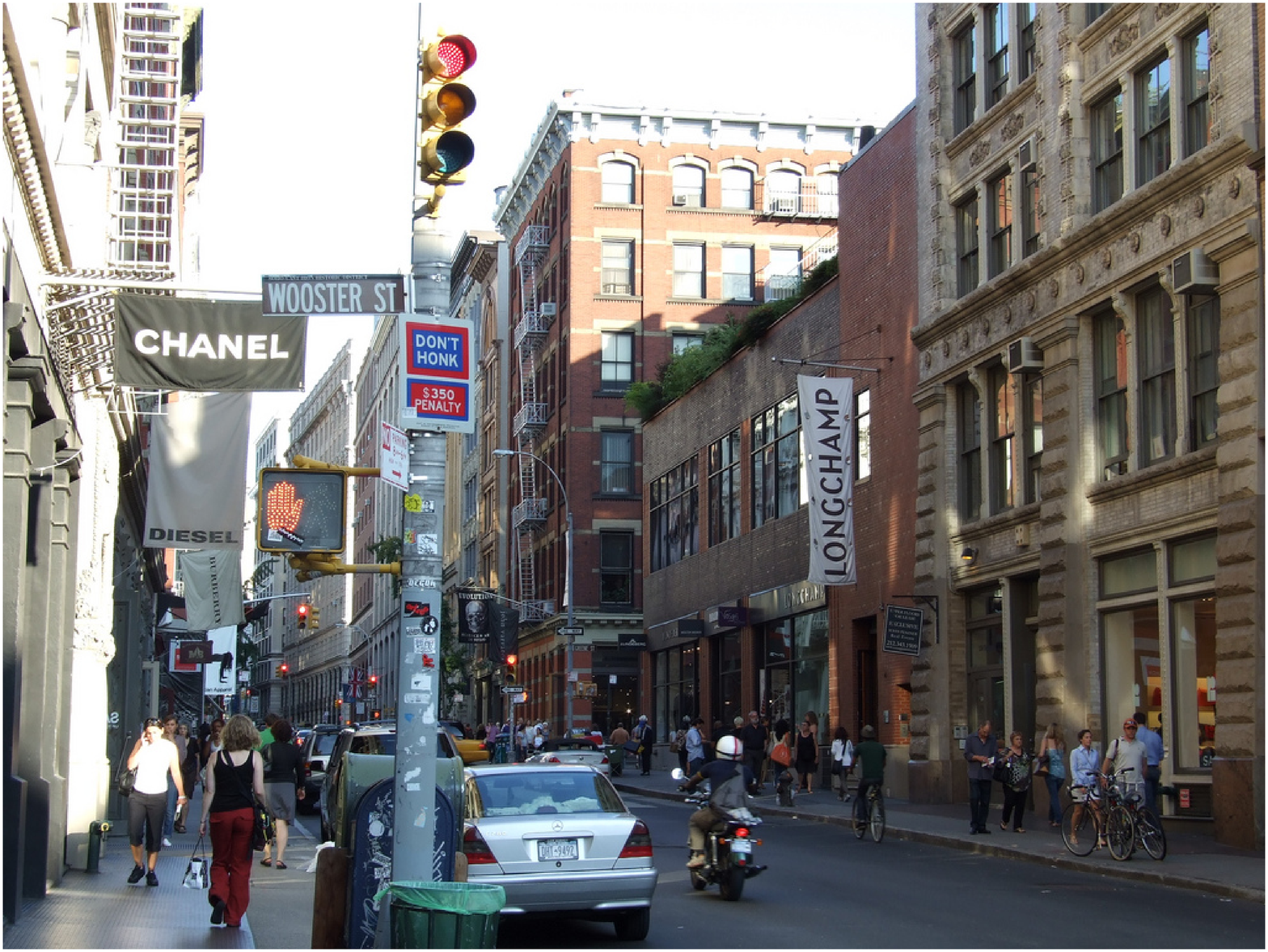, width= 6cm, height =4cm}
& 4). &  \epsfig{file= 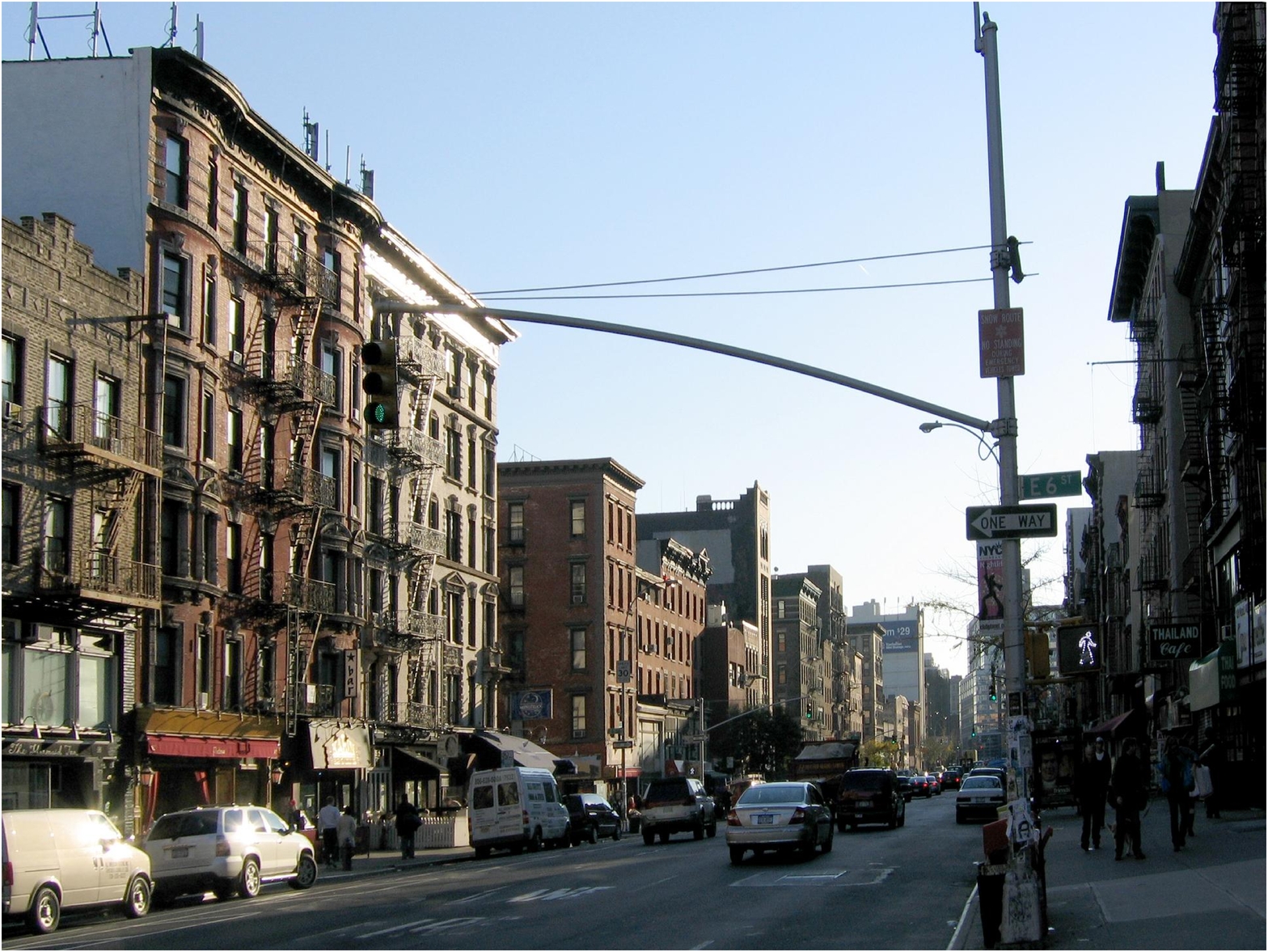, width=6cm, height =4cm} \\
5). & \epsfig{file= 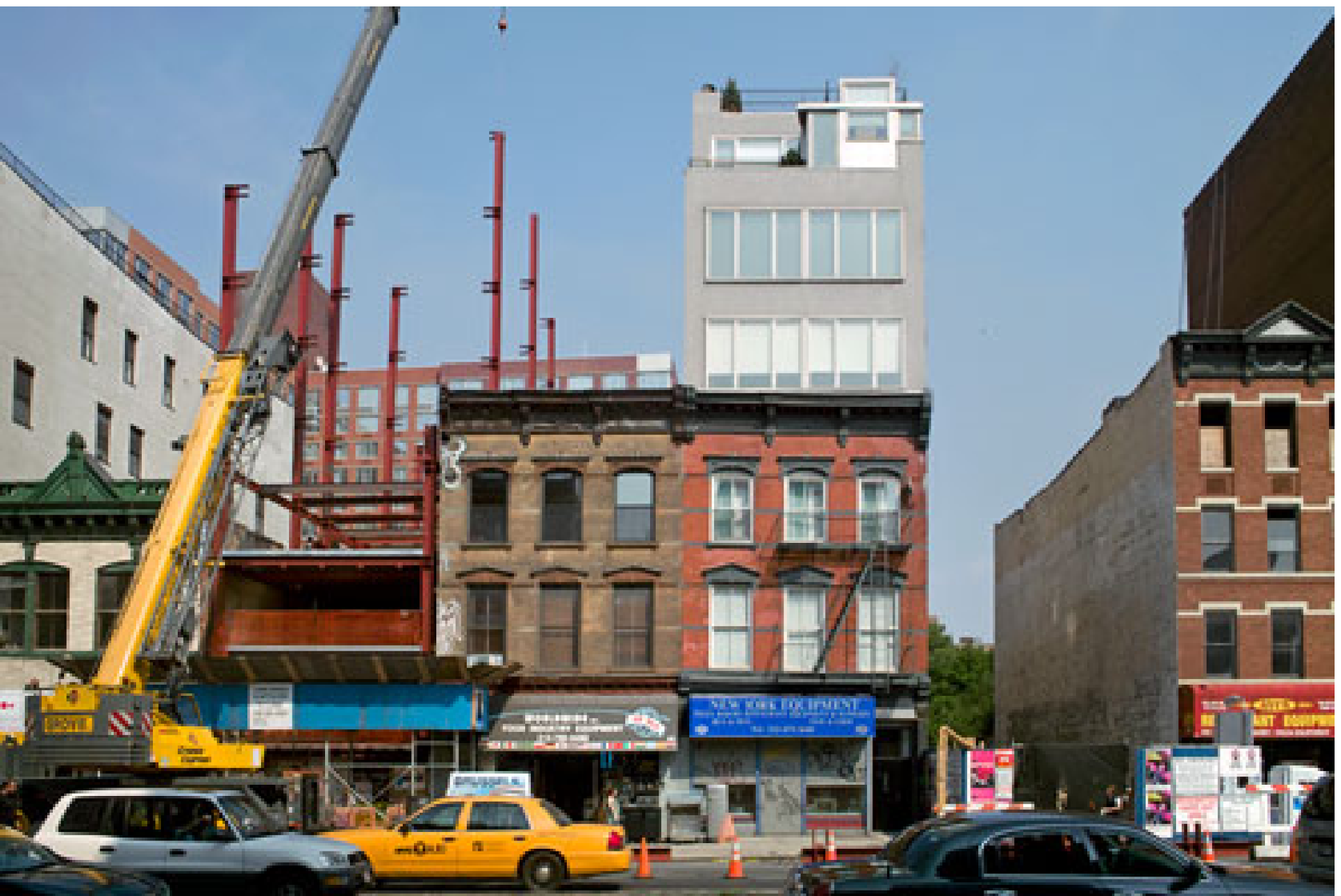, width= 6cm, height =4cm}
& 6). &  \epsfig{file= 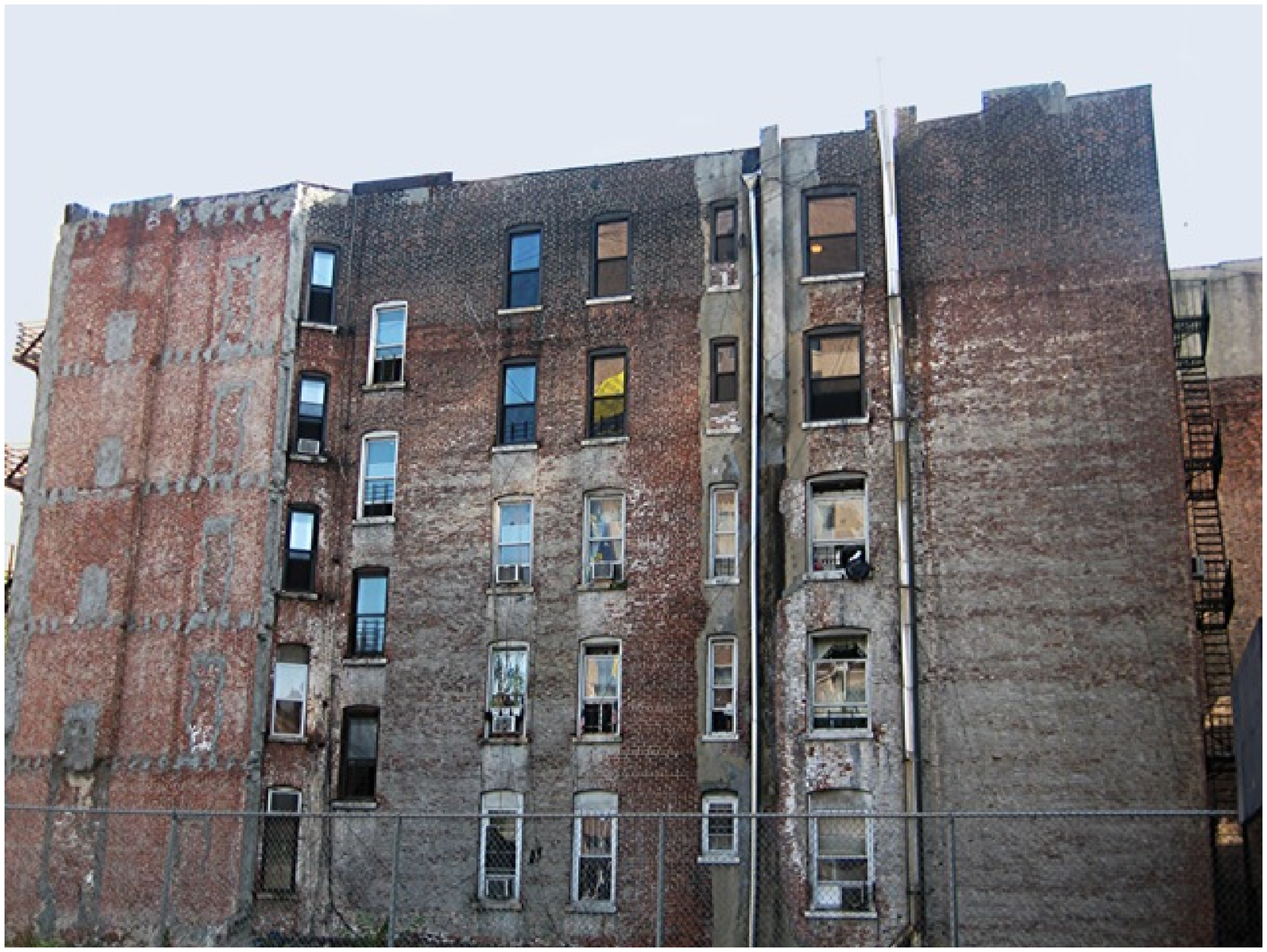, width=6cm, height =4cm} \\
\end{tabular}
\caption{ 
The first passage times in the borough of Manhattan, NYC: 
1). the Federal Hall National Memorial $\sim$ 10 steps; 
2). the Times square  $\sim$ 100 steps;
3). the SoHo neighborhood, in Lower Manhattan  $\sim$ 500 steps;
4). the East Village neighborhood, lying east of Greenwich Village, south of Gramercy and Stuyvesant Town  $\sim$ 1,000 steps;
5). the Bowery  neighborhood, in the southern portion of the New York City borough of Manhattan $\sim$ 5,000 steps;
6). the East Harlem (Spanish Harlem, El Barrio),  a section of Harlem located in the northeastern extremity of the borough of Manhattan $\sim$ 10,000 steps;
\label{Fig_Manhattan}}
\end{figure}

The  first-passage times enable us
to classify all places in the
 spatial graph of Manhattan
into four groups accordingly
to the first-passage times to them \cite{Blanchard:2009}.
The first group of locations
is characterized by the minimal first-passage times;
 they are probably reached for the first time
 from any other place
of the urban pattern  in just
 10 to 100 random navigational steps (the heart of the city), 
see Fig.~\ref{Fig_Manhattan}.1 and Fig.~\ref{Fig_Manhattan}.2.
These locations are identified as belonging to the downtown of
Manhattan (at the south and southwest tips of the island) -- the
Financial District and Midtown Manhattan. It is interesting to
note that these neighborhoods are roughly
coterminous with the
boundaries
 of the ancient New Amsterdam settlement founded in the
late 17${}^{\mathrm{th}}$ century.
Both districts comprise the offices and headquarters
of many of the city's major financial institutions
such as the New York Stock Exchange and the American Stock Exchange
(in the Financial District). Federal Hall National Memorial is also
encompassed in this area that had been
 anchored by the World Trade Center until the September 11, 2001
 terrorist attacks.
We might conclude that the group of locations characterized by the
best structural accessibility is the heart of the  {public process
in the city}.

The neighborhoods from the second
group (the  city core)
 comprise the locations
that can be reached for the first time
 in several hundreds
to roughly a thousand
 random navigational steps
from any other place of the urban pattern
(Fig.~\ref{Fig_Manhattan}.3 and Fig.~\ref{Fig_Manhattan}.4).
SoHo (to the south of Houston Street), Greenwich Village, Chelsea (Hell's Kitchen),
the Lower East Side, and the East Village
are among them --
they are commercial in nature and known for upscale shopping and
the "Bohemian" life-style of their dwellers
contributing into New York's art industry and nightlife.

The relatively isolated neighborhoods such
as Bowery (Fig.~\ref{Fig_Manhattan}.5), some segments in Hamilton Heights and Hudson Heights,
Manhattanville (bordered on the south by Morningside Heights), TriBeCa
(Triangle Below Canal)
and some others can be associated to
 the third structural category
as being reached for the first time from 1,000 to 3,000 random
steps starting from a randomly chosen place in the spatial graph
of Manhattan. 
Interestingly, that
many locations belonging to the third structural
group comprises the
diverse and eclectic mix of different
social and religious groups.
Many famous  houses of worship
had been established there during the late
19${}^{\mathrm{th}}$ century --
St. Mary's Protestant Episcopal Church,
Church of the Annunciation,
St. Joseph's Roman Catholic Church,
and Old Broadway Synagogue in Manhattanville
are among them. The  neighborhood of  Bowery
in the southern portion of  Manhattan
had been most often associated with the poor and the homeless.
From the early 20${}^{\mathrm{th}}$ century, Bowery
became the center of the so called "b'hoy"
subculture of working-class
 young men
 frequenting the cruder nightlife.
Petty crime and prostitution followed in their wake, and
most respectable businesses, the middle-class,
 and entertainment had fled the area.
Nowadays, the dramatic decline has lowered crime rates in the
district to a level not seen since the early 1960s and continue to
fall. Although zero-tolerance policy targeting petty criminals is
being held up as a major reason for the crime combat success, no
clear explanation for the crime rate fall has been found.

The last structural category comprises
 the most isolated segments in the city
mainly allocated in the Spanish and East Harlems.
They are characterized by
the longest first-passage times from 5,000 to 
10,000  random steps (Fig.~\ref{Fig_Manhattan}.6).
 Structural isolation is fostered
by  the
unfavorable confluence of many factors
 such as
the close proximity to Central Park
(an area of
340 hectares
removed from the otherwise regular street grid),
the
boundness by the strait of
 Harlem River
separating the Harlem and the Bronx,
and the remoteness from the main
bridges (the Triborough Bridge,
the Willis Avenue Bridge, and
the Queensboro Bridge) that
  connect the boroughs of Manhattan to
the urban arrays in
Long Island City and Astoria.
Many social problems associated with poverty
 from crime to drug addiction have plagued
the area for some time. The haphazard change of the racial
composition of the neighborhood occurred at the beginning of the
20${}^{\mathrm{th}}$ century together with the lack of adequate
urban infrastructure and services fomenting racial violence in
deprived communities and made the neighborhood unsafe -- Harlem
became a  {slum}. The neighborhood had suffered with unemployment,
poverty,  and crime for quite long time and even now, despite the
sweeping economic prosperity and redevelopment of many sections in
the district, the core of Harlem remains poor.

Recently, we have discussed in \cite{Blanchard:2011}
that distributions of
 various social variables
(such as the mean household income and prison expenditures in
different zip code areas) may demonstrate the striking spatial
patterns which can be analyzed by means of random walks. In the
present work, we analyze the spatial distribution of the tax
assessment rate (TAR) in Manhattan.

The assessment tax relies upon a special enhancement made up of
the land or site value and differs from the market value
estimating a relative wealth of the place  within the city
commonly referred to as the 'unearned' increment of land use,
\cite{Bolton:1922}. The rate of appreciation in value of land is
affected by a variety of conditions, for example it may depend
upon other property in the same locality, will be due to a
legitimate demand for a site, and for occupancy and height of a
building upon it.

 \begin{figure}[ht]
 \noindent
\begin{center}
\epsfig{file=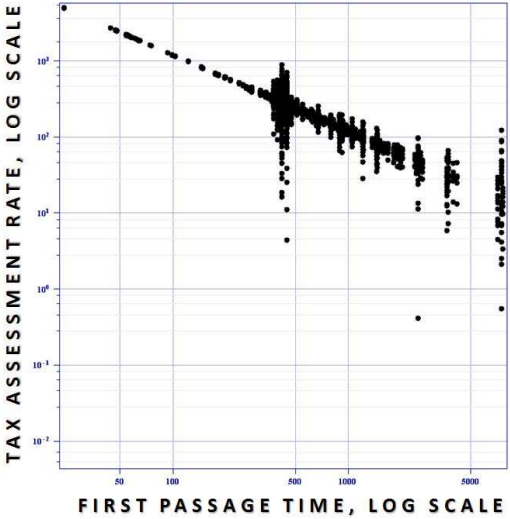, angle= 0,width =7cm, height =7cm}
  \end{center}
\caption{\small Tax assessment rate (TAR)  of places in Manhattan
 (the vertical axes, in \$/fit${}^{2}$)
 is shown in the logarithmic scale
vs.
the first--passage times (FPT) to them (the horizontal axes).  }
 \label{Fig2_prices}
\end{figure}

The current tax assessment system enacted in 1981
in the city of New York
 classifies all real estate parcels into four classes subjected
to the different tax rates set by the legislature:
(i) primarily residential condominiums; (ii) other residential property;
(iii) real estate of utility corporations and special franchise properties;
(iv) all other properties, such as stores, warehouses, hotels, etc.
However, the scarcity
of physical space in the compact urban pattern on the island of Manhattan
will naturally set some increase of value on all desirably located
land as being a restricted commodity.
Furthermore, regulatory constraints on housing supply exerted on housing prices
by the state and the
city in the form of 'zoning taxes'
are responsible for
converting the property tax system in a complicated mess
of interlocking influences and for much of the high cost of housing
in Manhattan, \cite{Glaeser:2003}. 

Being intrigued with the
 likeness of
the tax
assessment map and the map of isolation in Manhattan,
 we have mapped the TAR figures publicly available
through the Office of the Surveyor at
the Manhattan Business Center onto the
 data on first-passage times to the corresponding
 places.
The resulting plot is shown in Fig.~\ref{Fig2_prices}, in the
logarithmic scale. The data presented in Fig.~\ref{Fig2_prices}
positively relates the geographic accessibility of places in
Manhattan
 with their 'unearned increments'
estimated by means of the increasing burden of taxation.
The inverse linear pattern dominating the data
is best fitted by the simple hyperbolic relation between
the tax assessment rate (TAR)
and the value of first--passage time (FPT),
$
\mathrm{TAR}\approx{c}/{\mathrm{FPT}},
$ in which $c\simeq 120,000\$\times\mathrm{Step}/\mathrm{fit}^2$ is a fitting
constant.

\section{Mosque and church in dialog}
\label{subsec:Mosque_and_kirchen}
\noindent

Churches are buildings used  as religious places,
 in the Christian tradition.
In addition to being a place of worship, the churches in  Western
Europe were utilized by the community in other ways -- they could
serve as a meeting place for guilds. Typically, their location
were at a focus of a neighborhood, or a settlement.

 \begin{figure}[ht]
 \noindent
\begin{center}
\epsfig{file=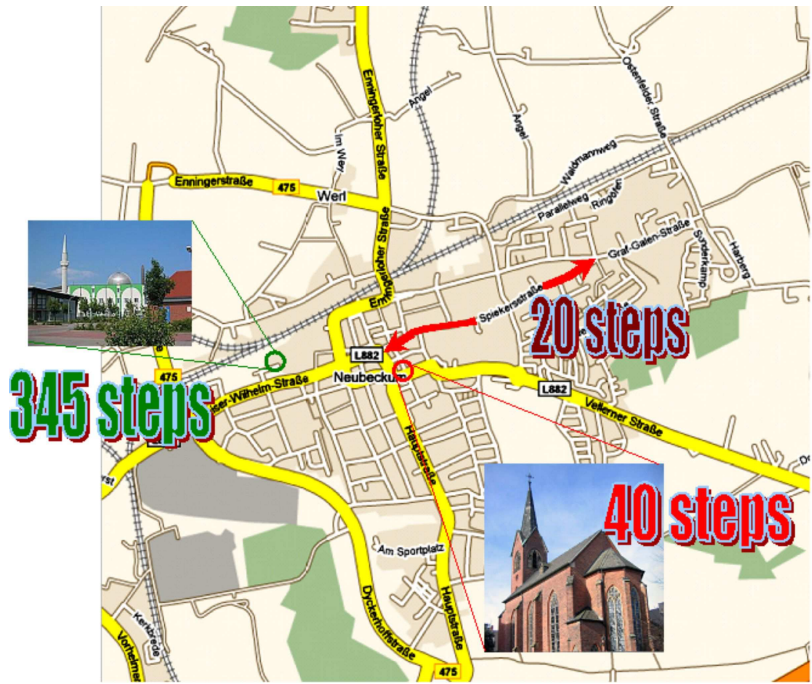, angle= 0,width =11cm, height =9cm}
  \end{center}
\caption{\small Neubeckum (Westphalia): the church and the mosque in dialog. }
 \label{Fig2_kirch}
\end{figure}

Nowadays, because of the intensive
 movement of people between countries,
the new national unities
 out of cultural and religious diversity
have appeared.
United States possessing
 rich tradition of immigrants
have demonstrated the ability of an increasingly multicultural
society to unite different religious, ethnic and linguistic groups
into the fabric of the country, and many European countries follow
that way, \cite{Portes:2006}.

Religious beliefs and institutions have played
 and continue to play a crucial role in new immigrant communities.
Religious congregations often provide ethnic,
cultural and linguistic reinforcements,
and often help newcomers to integrate
by offering a connection to social groups
that mediate between the individual and the new society,
so that
immigrants often become even more religious once in
the new country of residence, \cite{Kimon:2001}.

It is not a surprise that
the buildings belonging to
 {religious congregations
of newly arrived immigrants}
are usually located not at the centers
of cities
 in the host country -- the changes
 in function results in a change of location.
In the previous section,
we have discussed that religious organizations
of immigrants
in the urban pattern of Manhattan
have been usually founded in the
 relatively isolated
locations, apart from the city core,
like those in Manhattanville.
 We have seen that
the typical first-passage times to the "religious" places of
immigrant communities in Manhattan scale from 1,000 to 3,000
random steps. It is interesting to check this observation also for
the  religious congregation buildings of recent immigrants in
Western Europe.

Despite
the
mosque and the church
are located in close geographic proximity
in the
city of Neubeckum
(Nordrhein-Westfalen, Germany),
their
 locations
are dramatically different with respect to the entire
 city structure.
The analysis of the
spatial graph of the city of Neubeckum
by random walks
shows that while the church
is situated in a place belonging
to the city core, and
 just 40 random steps
are  required
  in order to reach it for the first time
 from
any arbitrary chosen place,
a random walker needs 345 random steps
to arrive at the mosque. The commute time, the
expected number of steps a random walker needs
to reach the mosque from the church and then to return
back, equals 405 steps.

Spiekersstrasse, the street which is parallel to the railway,
now is the best accessible place of motion in Neubeckum
playing the role of its structural "center of mass";
it can be achieved from any other location
in the city in just 20 random steps.
The relation between the extent
of structural isolation and the
specified reference levels
  can be measured
in a logarithmic scale by using as unit of  {decibel} (dB), \cite{Blanchard:2009}.
When referring
to estimates of isolation
 by means of first-passage times
(FPT),
 a ratio between two levels
inherent to the different locations A and B
can be expressed in decibels by evaluating,
$
\mathrm{I}_{AB} = 10\,\log_{10}\left( { \mathrm{FPT}(A) }/{\mathrm{FPT}(B)}\right),
$
where $\mathrm{FPT}(A)$ and $\mathrm{FPT}(B)$ are the first-passage times
to A and B respectively. 
If we estimate relative isolation of
other places of motion
with respect to  Spiekersstrasse
by comparing their first-passage times in the logarithmic scale,
then
the location of the church
is evaluated by
$I_{\mathrm{Church}}\approx 3\,$dB of isolation, and
$I_{\mathrm{Mosque}}\approx 12\,$dB, for the mosque.

Indeed, isolation was by no means the aim of the Muslim community.
The mosque in Neubeckum has been erected on a vacant place, where
land is relatively cheap. However,  structural isolation under
certain conditions would potentially have dramatic social
consequences. Efforts to develop systematic dialogue and increased
cooperation based on a reinforced culture of consultations are
viewed as essential to deliver a sustainable community.

\section{Which place is the ideal Bielefeld crime scene?}
\label{sec:Bielefeld}
\noindent

Bielefeld is a city in  the north-east of North-Rhine Westphalia (Germany)
famous as a home to a significant number of internationally operating companies.

{\it "Which place is the ideal Bielefeld crime scene?"}
This question has been recently addressed by the {\it Bielefeld--heute} ({\it "Bielefeld today"}) weekly newspaper to those crime fiction authors who had chosen Bielefeld as a stage for the criminal stories of their novels.
 Although the above question falls largely within the domain of criminal psychology, it can also be considered as a problem of mathematics -- since the limits of human perception coincide with mathematically plausible solutions. 

\begin{figure}[ht!]
 \noindent
\begin{center}
\epsfig{file=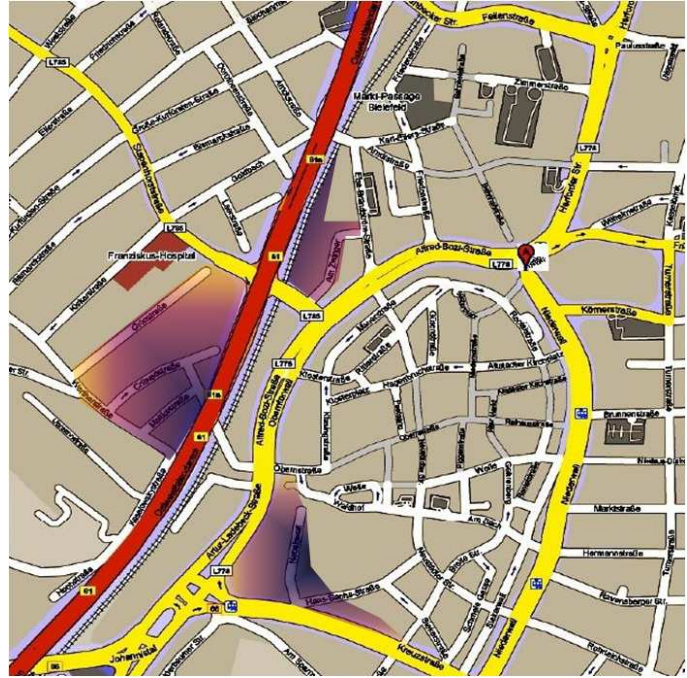,  width =9cm, height =9cm}
  \end{center}
\caption{\small  The most isolated places in the city of Bielfeld. }
 \label{Fig2_Biel}
\end{figure}

We have analyzed how easy it is to get to various places on the labyrinth in the network of 200 streets located at the city centre of Bielefeld aiming to capture a neighborhood's inaccessibility which could expose hidden islands of future deprivation and social misuse in that.  
For our calculations, we imagined pedestrians wandering randomly along the streets and worked out the average number of random turns at junctions they would take to reach any particular place in Bielefeld from various starting points. 
Not surprisingly, the {\it August-Bebel Str.}, {\it Dorotheenstra{\ss}e} and the {\it Herforder Str.} were the most accessible in the city. In contrast, we found that the certain districts located along the rail road (see the map shown in Fig.~\ref{Fig2_Biel}) jumped out as by far the most isolated, despite being apparently well connected to the rest of the city.

\begin{figure}[ht!]
 \noindent
\centering
\begin{tabular}{llll}
1). & \epsfig{file= 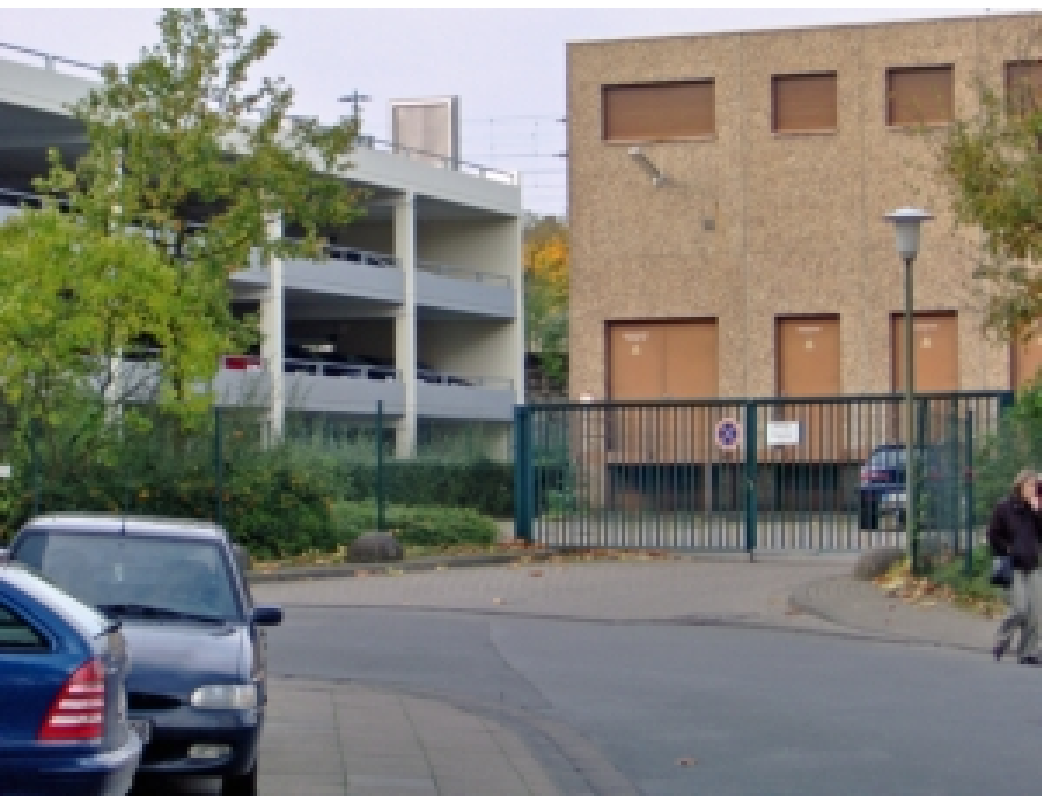, width=6cm, height =4cm}
& 2). &  \epsfig{file= 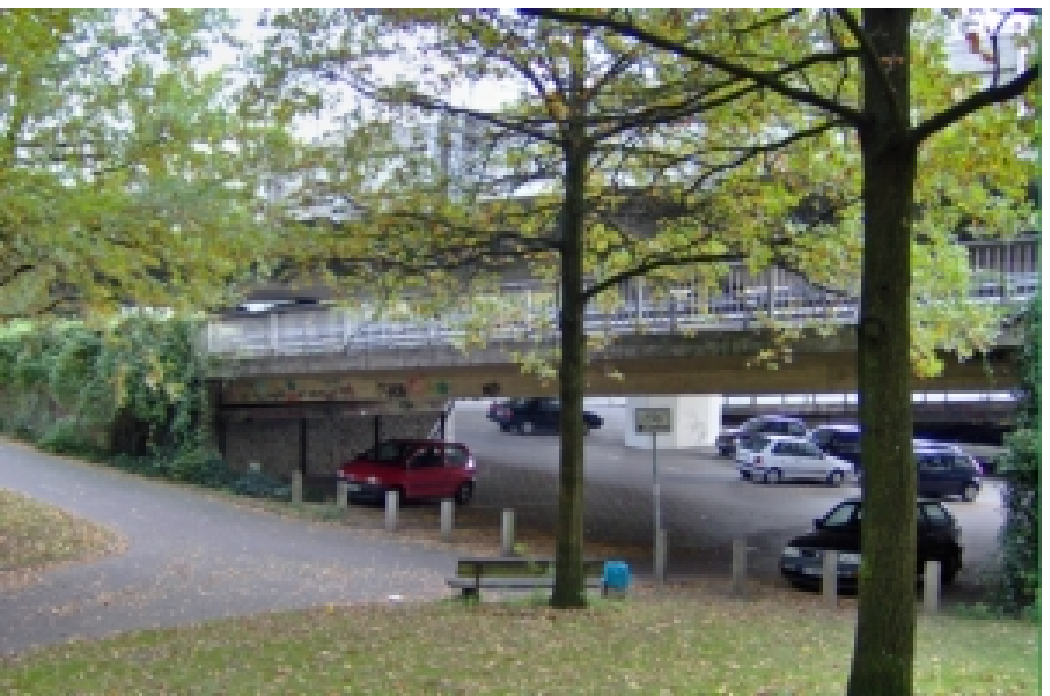, width=6cm, height =4cm} \\
\end{tabular}
\begin{tabular}{ll}
3). & \epsfig{file= 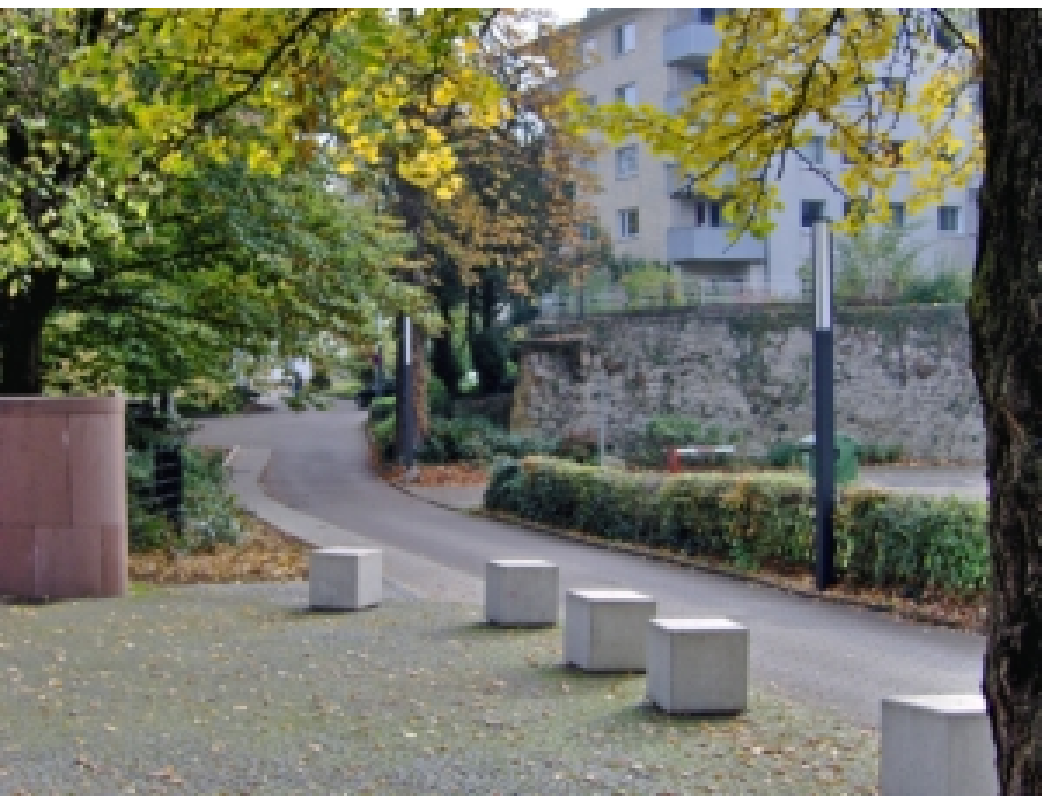, width=6cm, height =4cm}
\end{tabular}
\caption{ The  ideal Bielefeld crime scene: 1). the parking place {\it Am Zwinger}; 2). the intersection of {\it Cr\"{u}wellstra{\ss}e} and {\it Moltkestra{\ss}e}; 3). the place beyond the Natural History Museum and the city Art Gallery.
\label{Fig_Biel02}}
\end{figure}

On average, it took from 1,389 to 1,471 random treads to reach such the god-forsaken corners as the parking places on {\it Am Zwinger} (Fig.~\ref{Fig_Biel02}.1), the neighbourhood centered by the {\it Cr\"{u}wellstra{\ss}e} (Fig.~\ref{Fig_Biel02}.2), the waste places close to the Natural History Museum and the city Art Gallery (Fig.~\ref{Fig_Biel02}.3) - far more than the average of 450 steps for other places in Bielefeld.

The inhospitable isolation of these places can be estimated numerically, however people rather percept it intuitively. Although the actual criminal rate in Bielefeld appears to be relatively low, many pulp fiction authors found the city a suitable place for their criminal stories that indeed recalls us the sustained satirical Internet-Myth of the {\it Bielefeld Verschw\"{o}rung} (Bielefeld Conspiracy). In spite of all efforts to subsidize development and publicity for Bielefeld by the city council, it still has a solid reputation for obscurity.

Our analysis shows that the city of Bielefeld consists of three structurally different components loosely tied together by just a few principal routes. Being founded in 1214 by Hermann IV, the Count of Ravensberg, the compact city guarded a pass crossing the Teutoburger Forest. In 1847, the new Cologne-Minden railway had passed through Bielefeld establishing the new urban development apart from the historical core of the major city - the Bahnhofsviertel governed mostly by the linear structure of the rail road. Finally, during the industrial revolution, the modern city quarters had been constructed by the end of the 19th century. These city districts built in accordance with different development principles and in different historical epochs are strikingly dissimilar in structure. Walkers in our model were mostly confined in each city domain experiencing difficulty while alternating that. Not surprisingly, most Germans have a vague image of the city in their heads.  The threefold structure of Bielefeld would make the city center extremely vulnerable to proliferation of growth problems.

\section{Conclusion and Discussion}
\label{sec:Conclusion}
\noindent

We  assumed
that
spatial experience in humans
intervening in the city
may be organized in the form of
a universally acceptable network. 
We also assumed that
the frequently travelled routes
 are nothing else
but the "projective invariants" of the
given layout of streets and squares in the city --
the function of its geometrical configuration,
which remains invariant whatever
origin-destination route is considered.

Basing on these two assumptions, we have developed
 a method that allows to
capture a neighborhood's inaccessibility.
Any finite undirected graph  can be
interpreted as a discrete
time dynamical system with a finite
number of states.
The temporal evolution of such a
dynamical system
is described by a "dynamical law" that
maps vertices of the graph into other vertices
and can be interpreted as
the transition operator of random walks.
The level of accessibility of nodes and
subgraphs of undirected graphs
can be estimated precisely
 in connection with random walks
introduced on them.
We have applied this method to the structural analysis
 of different cities.

The main motivation of our work was to get an insight into the structure of human settlements that would improve the overall strategy of investments and planning and avoid the declining of cities as well as the many environmental problems.
Multiple increases in urban population
that had occurred in Europe
 at the beginning of the 20$^\mathrm{th}$ century
have been among the decisive factors
that
 changed the world.
Urban agglomerations had suffered from the  {co-morbid problems} such
as widespread poverty, high unemployment, and rapid changes in the
racial composition of neighborhoods. Riots and social revolutions
have occurred in urban places in many European countries in part
in response to deteriorated  conditions of urban decay and
fostered political regimes affecting immigrants and certain
population groups {\it de facto} alleviating the burden of
 the haphazard urbanization by increasing its deadly price:
tens of millions of people had emigrated from Europe,
but many more of them had died of starvation and epidemic diseases,
or became victims of wars and political repressions.

Urbanization has been the dominant demographic trend in the entire world, during the last half century. 
Although
the intense  process
of urbanization
is a proof of economic dynamism,
clogged roads, dirty air, and
deteriorating neighborhoods are
 fuelling a backlash
against urbanization that
nevertheless cannot be stopped.
The urban
design decisions made today on the base of the US car-centered
model, in cities of the developing world where car use is still
low, will have an enormous impact on climate changes in the decades
ahead. Unsustainable pressure on resources causes
 the increasing loss of fertile lands through
degradation and the dwindling amount of fresh
water and food would trigger conflicts and result
 in   {mass migrations}.
Migrations induce a dislocation and disconnection between the
population and their
 ability to undertake   {traditional land use}, \cite{Fisher:2008}.
Major metropolitan areas and the
intensively growing urban agglomerations
attract large numbers of   {immigrants with limited skills}.
Many of them will end up a burden on the state,
 and perhaps become involved in criminal activity. 
 The poor are urbanizing faster than the population as a whole, \cite{Ravallion:2007}.
  {Global poverty} is in flight becoming a primarily
 urban phenomenon in the developing world: about 70\% of 2 bln new urban settlers
in the next 30 years will live in slums,
 adding to 1 bln already there.
The essential attention should be given to the cities in the developing world where the accumulated urban growth will be duplicated in the next 25 years. 
The fastest urbanization of poverty
 occurred in Latin America, where the majority
of the poor now live in urban areas.

Faults in urban planning, poverty, redlining,
immigration restrictions and clustering of minorities
dispersed over the spatially isolated pockets of
 streets trigger urban decay, a process by which a
city falls into a state of disrepair.
The speed and scale of urban growth require urgent global actions
to help cities prepare for growth and to avoid them of being the
future epicenters of poverty and human suffering.

People of modern Europe prefer to live in single-family houses and commute by automobile to work. In 10 years (1990-2000), low-density expansions of urban areas known as 'urban sprawl' consumed more than 8,000 $\mathrm{km}^2$ in Europe, the entire territory of the state of Luxembourg. Residents of sprawling neighborhoods tend to emit more pollution per person and suffer more traffic fatalities. Faults in planning of urban sprawl neighborhoods would force the structural focus of the city out from its historical center and trigger the process of degradation in that. 

Together with severe environmental problems generated by the unlimited expansion of the city, the process of urban degradation creates dramatic economic and social implications, with negative effects on the urban economy. It is well known that degraded urban areas are less likely to attract investments, new enterprises and services, but become attractive for socially underprivileged groups because of a tendency of reduction house prices in the urban core. Smart growth policies that concentrate the future urban development in the center of the city to avoid urban sprawl should be applied.

Our  last but not least remark is that sprawling suburbs in USA saw by far the greatest growth in their poor population and by 2008 had become home to the largest share of the nation's poor. Between 2000 and 2008, sprawls in the US largest metro areas saw their poor population grow by 25 percent - almost five times faster than primary cities and well ahead of the growth seen in smaller metro areas and non-metropolitan communities. These trends are likely to continue in the wake of the latest downturn, given its toll on the faster pace of growth in suburban unemployment.

A combination of interrelated factors, including urban planning decisions, poverty, the development of freeways and railway lines, suburbanization, redlining, immigration restrictions would trigger urban decay, a process by which a city falls into a state of disrepair. 
We often think that we have much enough time on our hands, but do we? 
The need could not be more urgent and the time could not be more opportune, to act now to sustain our common future.

\section{Acknowledgement}
\noindent  

The authors gratefully acknowledge
the financial support by the project 
{\it MatheMACS}
 ("Mathematics of Multilevel Anticipatory Complex Systems"),
 grant agreement no. 318723,
funded by the EC Seventh Framework Programme
 FP7-ICT-2011-8.

\end{document}